\documentclass[twocolumn,preprintnumbers,amsmath,amssymb,prb,showpacs,superscriptaddress]{revtex4}
\usepackage{amssymb}

\usepackage{graphicx,calc}
\usepackage{bm}
\usepackage[colorlinks=true,dvipdfm]{hyperref}

\newcommand{\tabincell}[2]{\begin{tabular}{@{}#1@{}}#2\end{tabular}}

\begin{document}
\title{Density Gradient Corrected Embedded Atom Method}
\author{Gang Wu}
\email[electronic mail: ]{wugaxp@gmail.com}
\author {Gang Lu}
\email[electronic mail: ]{ganglu@csun.edu}
\address{Department of Physics and Astronomy,
California State University Northridge, Northridge, CA 91330-8268 }
\author{Carlos J. Garc\'{\i}a-Cervera}
\address{Mathematics Department, University of California Santa
Barbara, CA 93106}
\author{Weinan E}
\address{Department of Mathematics, Princeton University,
Princeton, NJ 08544-1000}

\begin{abstract}
Through detailed comparisons between Embedded Atom Method (EAM) and first-principles
calculations for Al, we find that
EAM tends to fail when there are large electron density gradients present.
We attribute the observed failures
to the violation of the uniform density approximation (UDA) underlying EAM.
To remedy the insufficiency of UDA, we propose a gradient-corrected EAM model
which introduces gradient corrections to the
embedding function in terms of exchange-correlation and kinetic energies.
Based on the perturbation theory of ``quasiatoms" and density functional theory,
the new embedding function captures the essential physics missing in UDA,
and paves the way for developing more transferable EAM potentials. With Voter-Chen
EAM potential as an example, we show that the gradient corrections can significantly improve the
transferability of the potential.
\end{abstract}

\pacs {
31.15.xv, 
61.50.Ah, 
62.20.F- 
}

\date{\today}
\maketitle

\section{Introduction}

Atomistic simulations have become an increasingly powerful tool in
materials research and a worthy partner of theory and experiment.
Among the great many atomistic models, the Embedded Atom Method
(EAM) \cite{eam, meam} has emerged as one of the most successful and
versatile approaches, representing the mainstay of empirical
atomistic simulations. To date, EAM has been applied to a variety of
material systems, such as liquids, metals and alloys,
semiconductors, ceramics, polymers, nano-structures, and composite
materials. Examples of problems that EAM has studied include
structure, energetics and dynamics of lattice defects,\cite{intro1,
intro2, intro3} elastic response and phonons,\cite{intro4, intro5,
intro6} fracture and plastic deformation,\cite{intro7, intro8,
intro9} surface and surface growth,\cite{intro10, intro12,
intro13, intro14} thermodynamics properties,\cite{intro15} and phase
transitions,\cite{intro16} etc. The applications of EAM simulations
have been reviewed in Ref. \onlinecite{intro17}. The success and
popularity of EAM are a consequence of its sound theoretical
foundation - the density functional theory (DFT) and its simple
analytical expression. The former assures that the
\textit{essential} physics be captured by EAM and the latter endows
EAM with excellent numerical efficiency, in par with pair
potentials.

Despite its great success, EAM suffers from a major deficiency - the
lack of transferability. Most of EAM models are only reliable in
regimes for which they were parameterized; beyond the regimes of
parametrization, the reliability of EAM potentials quickly
deteriorates. As a result, the predicability of EAM is often
questionable in defect systems and in non-equilibrium conditions
where relevant physical quantities are not known accurately
\textit{a priori} and hence not included in the parametrization of
the potentials. As to all empirical models, the lack of
transferability of EAM is an indication that some theoretical
approximations of EAM model are not generally valid.

In this paper, we show that the lack
of transferability of EAM is attributable to the uniform background
density approximation of EAM embedding function. We find that EAM
fails whenever there are large gradients of electron density in the
system. We overcome the deficiency of the uniform density
approximation (UDA) by proposing a density gradient corrected EAM
model which incorporates the gradients of the valence electronic
density in the embedding function. Specifically, we introduce
additional terms into the embedding function which correspond to the
density gradient corrections to the exchange-correlation and kinetic
energy contributions. Motivated by the Perdew-Burke-Ernzerhof (PBE)
\cite{pbe} Generalized Gradient Approximation (GGA) of DFT and the
perturbation theory of ``quasiatoms" \cite{quasiatom}, the present
model applies to an inhomogeneous background density and has the
correct limiting behavior as the exact energy functions. As a
consequence, it extends the applicability of EAM and paves the way
for developing more transferable potentials.

\section{Failures of the uniform density approximation}

First, we demonstrate that the failure of EAM can be linked
to the presence of large gradients of electron density by comparing EAM with
first-principles DFT calculations. We establish this fact in bulk Al for which
EAM is supposed to work very well. Several excellent EAM potentials \cite{ea, mfmp, vc}
exist and they are used for comparisons. Both
elastic properties and stacking fault energy of Al are calculated. We
compute the cohesive energy per atom and the stress tensor as a
function of the right Cauchy-Green deformation tensor $C_{ij}$ ($i,
j =1, 2, 3$) for a primitive unit cell of bulk Al. There are six independent $C$ elements,
with the diagonal and off-diagonal elements varying from -0.28 to
0.28 and -0.18 to 0.18, respectively, giving rise to different elastic deformations.
In order to provide precise comparisons, we utilize the sparse grid
method \cite{spg} - a novel algorithm that allows us to represent a fine
high-dimensional mesh very efficiently. Specifically, we sample 483,201 data points in
sparse grids which correspond to a regular grid with 65 points in each
of the six dimensions of $C_{ij}$ and $65^6 \approx 7.5\times10^{10}$ points
in total. Moreover, by taking advantage of the underlying symmetry of
the system, we further reduce the number of data points from 483,201
to 24,567 for which we carry out first-principles and EAM calculations.

The first-principles DFT calculations are based on the plane-wave
and Projector Augmented-Wave method \cite{paw} as implemented in the
Vienna \textit{Ab initio} Simulation Package (VASP) \cite{vasp1,
vasp2}. We use PBE-GGA with a high plane-wave cutoff energy of 360
eV to obtain reliable energy and stress. The $k$-points are sampled
according to Monkhorst-Pack method with the $k$-point spacing less
than 0.0252 \AA$^{ - 1}$. A Gaussian smearing of 0.1 eV is employed
to speed up the convergence of the calculations. As for EAM
calculations, we employ three widely used EAM potentials, developed
by Ercolessi and Adams \cite{ea}, Mishin et al. \cite{mfmp}, and
Voter and Chen \cite{vc}. The first two potentials were constructed
by fitting to both experimental and first-principles data, while
Voter-Chen potential was fitted only to experimental data. Although
some EAM potentials \cite{ea} were constructed with different
motivations from the original EAM model, they all use the UDA in
effect. For the convenience of presentation, we define $\Delta V =
\frac{V - V_0 }{V_0 }$ and $\Delta \Omega = \frac{\Omega - \Omega _0
}{\Omega _0 }$, where $V$ ($V_0$) and $\Omega$ ($\Omega_0$) are the
volume and solid angle of the deformed (undeformed) unit cell. The
solid angle is defined relative to the basis vectors of the unit
cell. $\Delta V$ and $\Delta \Omega$ characterize the volumetric and
non-volumetric deformation of the unit cell.

\begin{figure}[htbp]
\includegraphics[width=\linewidth]{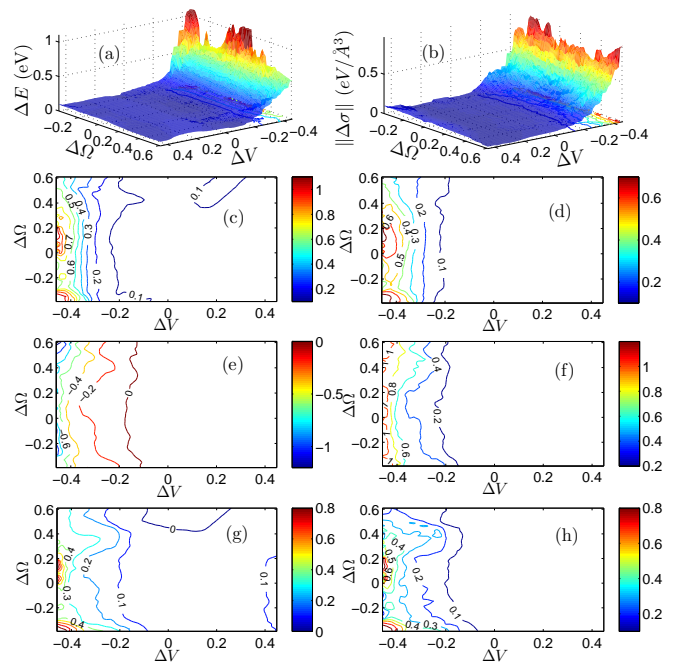}
\caption{The difference for the elastic properties calculated by EAM and VASP.
(a), (c), (e), and (g) show the cohesive energy differences in eV/atom;
(b), (d), (f), and (h) show the stress  differences in eV/\AA$^3$.
(a), (b), (c), and (d) are calculated by Ercolessi-Adams
potential. (c) and (d) are the corresponding contour plots of (a) and (b),
respectively. (e) and (f) are calculated by Mishin potential and (g) and (h) are
calculated by Voter-Chen potential. (e), (f), (g), and (h) are contour plots.}
\label{fig1}
\end{figure}

In \autoref{fig1}, we present the difference in the cohesive energy
$\left(\Delta E = E_{{\rm EAM}} - E_{{\rm VASP}} \right)$ and the
stress tensor $\left(\left\| {\Delta \sigma } \right\| = \left\|
{\sigma _{\rm EAM} - \sigma _{\rm VASP} } \right\|\right)$. Overall,
we find excellent agreement between the first-principles and EAM
results over a wide range of deformations - a remarkable feat of
EAM. Furthermore, the errors are insensitive to the solid angle,
$\Delta \Omega$, which reflects the delocalized nature of the
metallic bonds in Al, and thus justify the use of an angular - independent 
model in Al. On the other hand, we find that
the EAM errors depend very sensitively on the change of volume,
$\Delta V$; in particular, the EAM values deviate significantly from
the first-principles results for large compressions. The errors in
energy can reach as high as 1 eV/atom for 40\% compression. This
dramatic difference cannot be accounted for by the fitting errors of
EAM because all three potentials show exactly the same behavior.
Moreover, the shortest interatomic distance in the compressed unit
cell is 2.1 \AA, which is still within the fitting range of the
potentials. For example, the fitting range of bond length is from
2.0 {\AA} to 6.3 {\AA} in Mishin potential and 2.0 {\AA} to 5.6
{\AA} in Ercolessi-Adams potential. The results suggest that the
errors come from the model itself.

When the interatomic distance decreases, the gradient of electronic
density increases. For a large compression, the electron density
gradient could become too large for the UDA of EAM to be valid.
Indeed, we find that the density gradient increases considerably in
compressions, with the maximum value of the gradients
rising from 0.38 \AA$^{-4}$ for the perfect lattice to 0.54
\AA$^{-4}$ for 40\% compression. On the other hand, the expansion of
the lattice reduces the density gradient, and thus does not violate
the UDA.



To further the argument, we perform additional calculations for the
generalized stacking fault energy ($\gamma $) surface, which along
with elastic constants, determines the plastic behavior of
materials. We have carried out 293 energy calculations for the
entire $\gamma$-surface with the sparse grid representation. The
supercell consists of 9 layers in the $\left\langle {111}
\right\rangle $ direction for both EAM and VASP calculations.

\begin{figure}[htbp]
\hypertarget{fig2ab}{
\includegraphics[width=\linewidth]{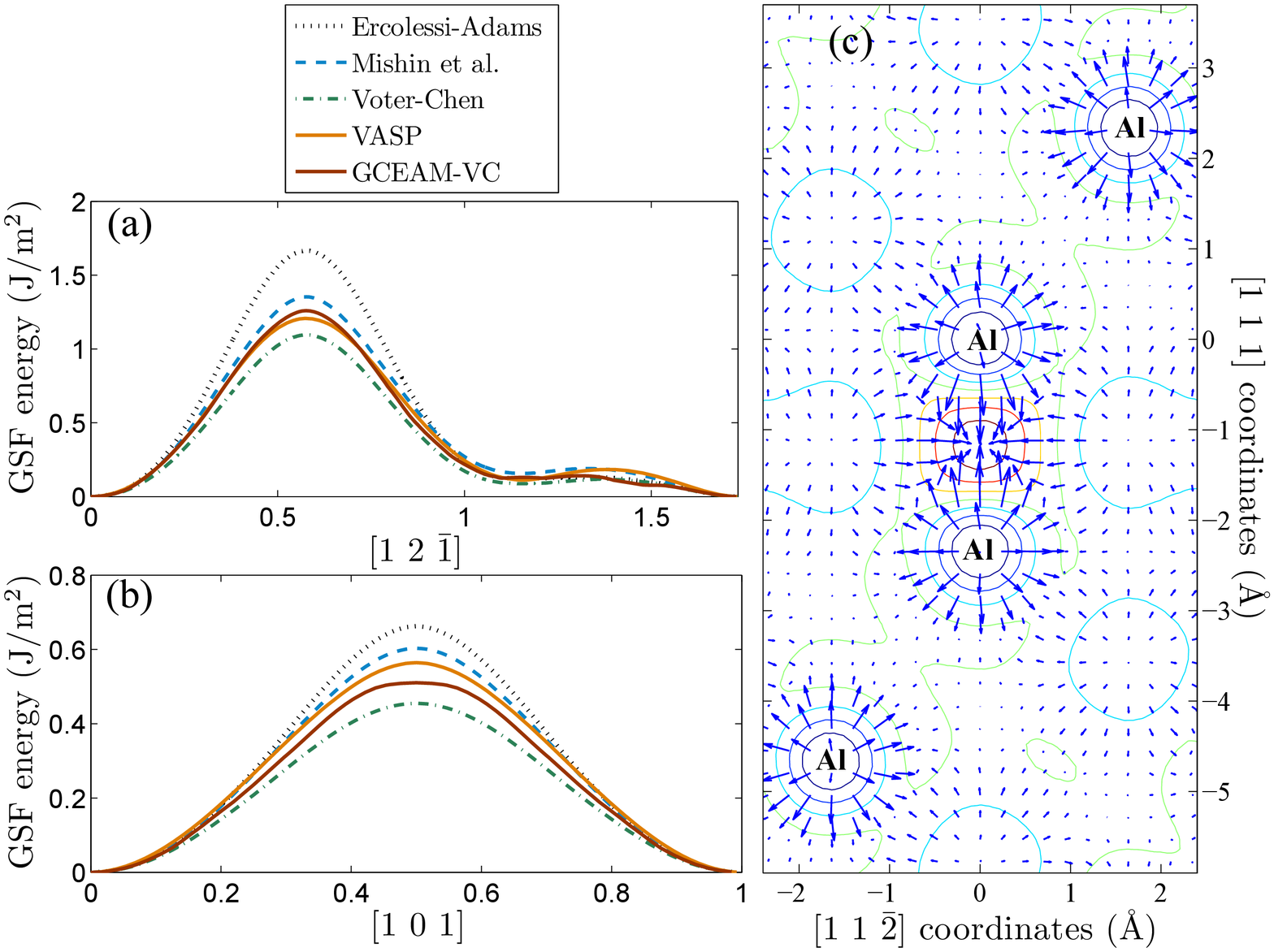}
}
\hypertarget{fig2c}{
\caption{$\gamma$-energy along (a) $\left[1 2 \bar{1}\right]$
and (b) $\left[1 0 1\right]$ directions from VASP and EAM calculations.
The horizontal axis is in the unit of the Burgers vector of Al (2.86 \AA).
The contours of valence electronic density and the density
gradient for the ``run-on'' configuration are shown in (c). The
arrow represents the direction and magnitude
of the density gradient.}}
\end{figure}

The $\gamma$-energy along $[1 2 \bar{1}]$ and $[1 0 1]$ directions
is shown in \hyperlink{fig2ab}{Figure 2a and 2b}. Again, overall
agreement between the three EAM potentials and VASP is good.
However, in the neighborhood of the [101] unstable stacking fault
and the run-on stacking fault (the last two entries in
\autoref{table1}), the magnitude of the energy error is significant.
In particular, the largest error of EAM occurs at the run-on
stacking fault in which the atoms in the two neighboring $\left(1 1
1\right)$ layers are right on top of each other, resulting in large
density gradients. The valence electronic density and its gradient
in the ``run-on'' configuration is presented in
\hyperlink{fig2c}{Figure 2c}. Noted that the maximum gradient of valence
electron density, $\left| g \right|_{\max }$
of both the unstable and the run-on stacking faults is comparable to
the corresponding value of large compressions ($\sim 0.5 $
\AA$^{-4}$), suggesting that the failures of EAM can be indeed
attributed to large density gradients, irrespective of the specific
atomic configurations. Furthermore, the data in the brackets of the
Table shows a general trend that the magnitude of EAM errors
increases as the maximum density gradient increases. Again the
shortest interatomic distance in \autoref{table1} is 2.33 \AA, which
is within the fitting range of bond length for the EAM potentials.

\begin{table}[htbp]
\caption{\label{table1} Fault vectors and energies for four stacking
faults obtained from VASP and EAM calculations. All
energies are in mJ/m$^{2}$. $\left| g \right|_{\max } $ denotes
the maximum gradient of valence electronic density calculated by
VASP, and is in \AA$^{ - 4}$. PBE, EA, Mishin and Voter
stand for the results calculated by VASP, Ercolessi-Adams,
Mishin et al., and Voter-Chen EAM potential, respectively. The errors
of EAM results are presented in brackets. $R_{\rm min}$ (\AA) is the
nearest neighbor distance in the corresponding configurations.}
\begin{ruledtabular}
\begin{tabular}{ccccccc}
Vector&
PBE&
EA&
Mishin&
Voter&
$\left| g \right|_{\max } $ &
$R_{\rm min}$\\
\hline
1/6[12$\bar{1}$]&
111&
121(10)&
157(46)&
87(-24)&
0.398 &
2.86\\
1/10[12$\bar{1}$]&
184&
132(-56)&
190(6)&
118(-66)&
0.399 &
2.74\\
1/4[101]&
564&
663(99)&
603(39)&
455(-109)&
0.507 &
2.47\\
1/3[12$\bar{1}$]&
1208&
1667(459)&
1354(146)&
1096(-112)&
0.581 &
2.33\\
\end{tabular}
\end{ruledtabular}
\end{table}

\section{Density gradient correction model}

Having established the importance of the density gradient, we
propose a gradient corrected model which could potentially improve
the transferability of EAM. The model is based on the pioneering
work of Stott and Zaremba \cite{quasiatom} on ``quasiatoms". Stott
\textit{et al.} have shown that by using a perturbation expansion
for an inhomogeneous background density, the embedding energy of a
``quasiatom" can be expressed rigorously as a function of the
\textit{background} density and its gradient. Based on the ``quasiatoms''
theory, we introduce three additional terms which account for the
gradient corrections to the exchange, correlation and kinetic energy
contributions to the embedding energy of EAM. In this context, the
original embedding function of EAM can be regarded as the UDA to the
embedding energy. Specifically, the corrected embedding function
becomes

\begin{eqnarray}
F_i\left( {\bar{\rho}_i,s_i} \right) &=& F_0 \left( \bar{\rho}_i
\right) + \tilde{F}_C \left( \bar{\rho}_i \right)g\left( s_i
\right)\nonumber\\
&{}& + \tilde{F}_X \left( \bar{\rho}_i \right)h\left( s_i \right) +
\tilde{F}_G \left( \bar{\rho}_i, s_i \right), \label{eq1}
\end{eqnarray}

\noindent where $\bar{\rho}_i \equiv \sum\limits_{j\neq i}
{\rho_j^{\rm at} (R_{ij})}$ is the background density at atom $i$,
and $\rho^{\rm at}_j$ is the density contribution from atom $j$.
$R_{ij}=\left| \vec{R}_i-\vec{R}_j \right|$ and $\vec{R}_i$ stands
for the atomic coordinates. Although the background density
$\bar{\rho}_i$ is not the same as the total density $\rho(\vec{r})$
where $\rho\left(\vec{r}\right)=\sum\limits_{j} {\rho_j^{\rm at}
\left(\left|\vec{r}-\vec{R}_j\right|\right)}$, they are closely
related and the gradients of both densities are well defined. In
particular, we can define a dimensionless background density
gradient $s_i$ as $s_i \propto {\left|\nabla_{\vec{R}_i}
\bar{\rho}_i \right|}/{\bar{\rho}_i^{4 / 3}}$, where
$\left|\nabla_{\vec{R}_i} \bar{\rho}_i\right|$ is the amplitude of
background density gradient. In practice, $s_i$ can be approximated
by its local average: $s_i \simeq \langle s_i \rangle \propto
\frac{1}{\left[ \bar{\rho}_i\right]^{4/3}} {\sum \limits_{j\neq i}
\left| {\frac{\partial \rho_j^{\rm at} \left( {R_{ij} }
\right)}{\partial R_{ij} }}\right|}$. $F_0 \left( {\bar \rho}
\right)$ is the UDA embedding function and $\tilde{F}_G$ is the
gradient correction to the kinetic energy. The leading term of
$\tilde{F}_G$ is of the von Weizs\"{a}cker form \cite{r14}, and can
be approximated as $\tilde{F}_G \left( \bar{\rho}, s \right) =
\tilde{K}_0 \left( \bar{\rho} \right) k\left( s \right)$. Here
$\tilde{K}_0$ resembles the Thomas-Fermi kinetic energy \cite{tf1,
tf2} and $k\left( s \right) = \lambda_0 \frac{1 + k_{11} s^2 +
k_{12} s^4}{1 + k_{21} s^2 + k_{22} s^4} s^2$. $\lambda_0$,
$k_{11}$, $k_{12}$, $k_{21}$ and $k_{22}$ are undetermined
parameters.

For exchange and correlation energy corrections,  we adopt the
functional form of PBE-GGA due to its simplicity. $\tilde{F}_C$ and
$\tilde{F}_X$ in \autoref{eq1} corresponds to the correlation and
exchange energy of the local density approximation (LDA) of DFT and
$g\left( s_i\right)$ and $h\left( s_i \right)$ are the corresponding
gradient corrections. The explicit forms $\tilde{F}_C$ and
$\tilde{F}_X$ can be found in standard references of LDA
\cite{pz,pw}. We assume spin degeneracy here although the spin
polarization can be considered easily and could be useful in the
development of spin-dependent EAM potentials for magnetic materials.
In addition, we require that the modified embedding functions have
the same limiting behavior as the exact functions:
\begin{equation}
\begin{aligned}
\left[g(s) \tilde{F}_C + h(s) \tilde{F}_X\right]
\begin{cases}
{\propto s^2 \tilde{F}_X,\hfill}&{s \to 0}\\
{\to - \tilde{F}_C + \kappa _0 \tilde{F}_X,\hfill}&{s \to \infty }
\end{cases}.
\end{aligned}
\label{eq2}
\end{equation}
We choose $g\left( s\right) = -\frac{{s}^4 }{g_0 + g_1 {s}^2 +
{s}^4}$ and $h\left( s \right) = \frac{\kappa_0 {s}^2 }{1 + \alpha
{s}^2}$, which satisfy the above conditions although other forms
of $g\left( s \right)$ and $h\left( s \right)$ can also be used.

Over all, there are six functions,
$\tilde{F}_C$, $\tilde{F}_X$, $\tilde{K}_0$, $\rho^{\rm at} \left( R
\right)$, $F_0 \left( \rho \right)$, and $\varphi \left(R\right)$, which are to be fitted.
The first three are new terms and in conjunction with
$g\left( s \right)$, $h\left( s \right)$ and $k\left( s \right)$,
they represent the gradient corrections to the embedding function.

Since the UDA functions, $\tilde {K}_0
\left( \bar {\rho } \right)$, $\tilde {F}_C \left( \bar {\rho }
\right)$, and $\tilde {F}_X \left( \bar {\rho } \right)$, have the
same functional forms as their DFT/LDA counterparts\cite{tf1, tf2, pz,
pw}, we have
\begin{subequations}
\label{ap-eq2}
\begin{equation}
\tilde {K}_0 \left( \bar {\rho } \right) \propto \bar {\rho }^{5 /
3}.
\end{equation}
\begin{equation}
\tilde {F}_C \left( \bar {\rho } \right) = \bar {\rho }\left( {c_1 +
c_2 r_s } \right)\ln \left( {1 + \frac{1}{\beta r_s^{p + 1} }}
\right), \quad r_s \propto \frac{1}{\bar {\rho }^{1 / 3}},
\end{equation}
\begin{equation}
\tilde {F}_X \left( \bar {\rho } \right) \propto \bar {\rho }^{4 /
3},
\end{equation}
\end{subequations}

Replacing the proportional sign `$ \propto $' in all the above
equations with an equal sign `=', one can fit the gradient corrected EAM (GCEAM)
potential by introducing 13 additional parameters. These additional parameters
are $c_1$, $c_2$, $\beta$, $p$, $g_0$, $g_1$, $h_0$, $h_1$,
$\lambda_0$, $k_{11}$, $k_{12}$, $k_{21}$, and $k_{22}$. Among them,
seven parameters ($k_{11}$, $k_{12}$, $k_{21}$, $k_{22}$, $g_0$,
$g_1$ and $\alpha$) are introduced through the gradient corrections.
$\kappa_0$ and $\lambda_0$ can be absorbed into the embedding
functions.


\section{Example: Gradient Corrected Voter-Chen potential}

In this section, we apply the gradient corrections to the Voter-Chen
(VC) potential, and the resultant potential is termed as GCEAM-VC.
It is important to mention that the corrections are not constrained
in any way by the specific form of the EAM potential. We choose the
VC potential because its simplicity - it has only five parameters,
much fewer than other EAM potentials, such as Mishin and
Ercolessi-Adams potentials. As a result, VC is not as accurate
as the other EAM potentials. However, the simplicity of the VC potential
renders more transparent physics, and frees us from intensive
parameter fitting  - which is not the emphasis of the present paper.
The goal of the article is to illustrate the importance of the
density gradient corrections in improving the transferability of EAM, rather than to generate the best possible EAM potential
for Al. Had we started from an EAM potential with more parameters, we would
have gotten even better test results for Al. Nevertheless, even with the VC
potential, the gradient corrections can significantly improve
the self-interstitial energies, stacking fault energies, etc. which
involve high density gradient configurations.

The GCEAM-VC potential takes the general form of EAM model.
The cohesive energy of a system can be written as

\begin{equation}
E = \frac{1}{2}\sum\limits_{i,j\left( { \ne i} \right)} {\varphi
_{ij} \left( {R_{ij} } \right) + \sum\limits_i {F_i \left( {\bar
{\rho }_i ,s_i } \right)} }, \label{ap-eq1}
\end{equation}

\noindent where embedding function $F_i\left( {\bar {\rho }_i ,s_i }
\right)$ is expressed in Eq. (\ref{eq1}).

The parameter fitting in GCEAM-VC follows the same procedure of the
Voter-Chen potential\cite{vc}, but with two modifications. The first
modification is that the pairwise interaction now is taken the form
of:

\begin{eqnarray}
\varphi \left( R \right) &=& \varphi _1 \left( R \right) + \varphi
_2
\left( R \right),\\
\varphi _1 \left( R \right) &=& D_M \left[ {1 - e^{ - \alpha _M
\left(
{R - R_M } \right)}} \right]^2 - D_M ,\nonumber \\
\varphi _2 \left( R \right) &=& C\left( {\frac{R_M }{R}} \right)^n.
\nonumber \label{ap-eq3}
\end{eqnarray}

Here, $\varphi _1 \left( R \right)$ is a Morse potential used in the
original Voter-Chen potential. $\varphi _2 \left( R \right)$ is
added to account for the repulsive interaction at short distance,
$\varphi \left( {R \to 0} \right) \to \infty $ when $C > 0$.
However, if one prefers to use fewer parameters, $\varphi _2 \left(
R \right)$ can be ignored without worsening the results (see the
discussions of Fig. \ref{ap-fig2}).

The second modification is that we do not fit the diatomic molecular
data. Instead, the force constants of bulk Al with different lattice
constants are fitted because accurate force constants give rise to
accurate phonon dispersions, and hence, accurate thermal properties
such as thermal capacity and conductivity. The force
constants are fitted for several lattice constants, including
0.9$a_0 $, 0.95$a_0 $, $a_0 $, 1.05$a_0 $, and 1.1$a_0 $, where $a_0
$ is the equilibrium lattice constant of bulk Al. It is found the
GCEAM-VC potential gives good description for the diatomic
properties without fitting them (See Table II). This is an example
of improved transferability of the GCEAM model.

\begin{figure}[htbp]
\includegraphics[width=0.92\linewidth]{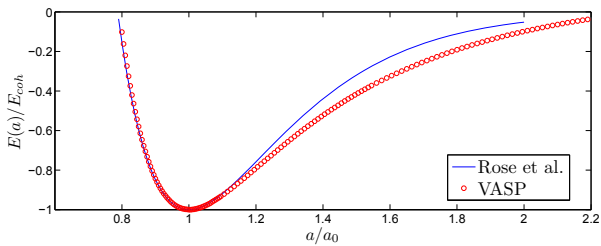}
\caption{The cohesive energy per atom of fcc Al as a function of the
lattice parameter (the scaled equation of states). $E_{coh}$ is the
cohesive energy for equilibrium lattice constant $a_0$. The Rose
\textit{et al.}'s universal EOS is represented by the blue line, and
the VASP values are represented by red open circles.}
\label{ap-fig1}
\end{figure}

\begin{figure}[htbp]
\includegraphics[width=0.92\linewidth]{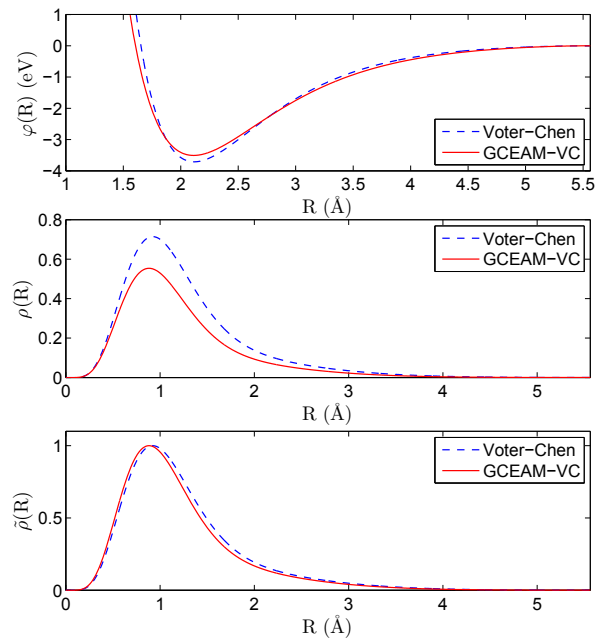}
\caption{The pair interaction function $\varphi(R)$ and atomic
density $\rho(R)$ of VC (blue dash) and GCEAM-VC (red solid)
potentials. } \label{ap-fig2}
\end{figure}

The embedding function of both VC and GCEAM-VC potentials is
determined by fitting the equation of states (EOS) to the universal
EOS of Rose \textit{et al.} \cite{EOS}. Because the universal EOS
does not agree exactly with the DFT (VASP)
values (see Fig. \ref{ap-fig1}), it is inevitable that both VC and
GCEAM-VC potentials would deviate from DFT results for large lattice
expansions or large interatomic distances. However, as shown later,
the gradient corrections can improve significantly the description of
high density gradient configurations involving lattice defects.

The density function of the VC potential is given as
\begin{equation}
\rho \left( r \right) = r^6\left[ {e^{ - \beta _2 r} + 2^9e^{ -
2\beta _2 r}} \right],
\end{equation}
\noindent $\beta_2$ needs to be fitted. We keep the same smoothness
conditions for the pairwise interaction, atomic density and EOS
function in GCEAM-VC as in the VC potential \cite{vc}, with the
cutoff radius $R_{\rm{cut}} $ of these functions to be fitted.
Thus, there are seven parameters, $D_M $, $\alpha _M $, $R_M
$, $C$, $n$, $\beta _2$ and $R_{\rm{cut}} $ to be fitted before
applying the gradient corrections. Overall, there are twenty
parameters in GCEAM-VC potential, including thirteen parameters
associated with the gradient correction terms. The optimized values
of all the parameters are presented in \autoref{ap-tab1}.

\begin{table}[htbp]
\caption{One set of optimized parameters for GCEAM-VC. Length and
energy unit are in $\AA$ and eV, respectively.}
\begin{ruledtabular}
\begin{tabular}
{lrlrlrlr} $c_1 $& -0.45618 & $g_1 $& -1.60375 & $k_{12} $&
-58.09505 & $R_M $&
1.34426  \\
$c_2 $& -1.03903 & $h_0 $& 0.41370 & $k_{21} $& -10.15397 & $C$&
0.43750  \\
$\beta $& 1.17295 & $h_1 $& 0.00327 & $k_{22} $& 26.82479 & $n$&
4.06038 \\
$p$& 0.29559 & $\lambda _0 $& -0.52795 & $D_M $& 4.00963 & $\beta _2
$&
3.46742 \\
$g_0 $& 1.63667 & $k_{11} $& 157.54365& $\alpha _M $& 2.05403 &
$R_{\rm{cut}} $&
5.56250  \\
\end{tabular}
\label{ap-tab1}
\end{ruledtabular}
\end{table}

The pair interaction $\varphi(R)$ and the atomic density $\rho(R)$
of GCEAM-VC potential are plotted in Fig. \ref{ap-fig2} in
comparison with the VC potential. It is found that the pair interaction
of GCEAM-VC changes very little from that of VC. Although the atomic
density function of GCEAM-VC potential appears to be rather
different from that of VC, this turns out not to be the case. Using
the fact that Eq. \ref{ap-eq1} is invariant under the
transformation:
\[
\rho \left( R \right) \to t\rho \left( R \right), F\left( \rho, s
\right) \to F\left( \rho/t, s/t \right),
\]

\noindent we can define a scaled atomic density $\tilde {\rho}
\left( R \right) = \rho \left( R \right)/ \max \left(\rho \left( R
\right)\right)$. $\tilde {\rho} \left( R \right)$ is plotted in the
bottom panel of Fig. \ref{ap-fig2} and one finds little difference
between the GCEAM-VC and VC atomic density functions. Therefore, we
conclude that all improvements to the VC potential come from
the gradient corrections, i.e., the model itself.

\begin{table}[htbp]
\caption{Properties of Al predicted by VC and and GCEAM-VC
potentials in comparison with experimental and/or \textit{ab initio}
data. $^*$Fitted properties.}
\begin{ruledtabular}
\begin{tabular}
{lccc} & \tabincell{c}{Experimental\\ or \textit{ab initio}}
&\tabincell{c}{Voter-Chen\\ (Ref. \onlinecite{vc})}&
GCEAM-VC \\
\hline Lattice properties:& & &
 \\
$a_0 $({\AA})$^*$& 4.05\footnotemark[1]& 4.05&
4.05 \\
$E_0 $(eV/atom)$^*$& -3.36\footnotemark[2]& -3.36&
-3.36 \\
$B$(GPa)$^*$& 79\footnotemark[3]& 79&
79.5 \\
$c_{11} $(GPa)$^*$& 114\footnotemark[3]& 107&
113 \\
$c_{12} $(GPa)$^*$& 61.9\footnotemark[3]& 65.2&
62.5 \\
$c_{44} $(GPa)$^*$& 31.6\footnotemark[3]& 32.2&
32.9 \\
Diatomic Properties:& & &
 \\
$D_e $(eV)& 1.60\footnotemark[4]& 1.54&
1.61 \\
$R_e $(\AA)& 2.47\footnotemark[4]& 2.45&
2.52 \\
Phonon frequencies:& & &
 \\
$\nu _L \left( X \right)$(THz)& 9.69\footnotemark[5]& 8.55&
9.62 \\
$\nu _T \left( X \right)$(THz)& 5.80\footnotemark[5]& 5.20&
5.36 \\
$\nu _L \left( L \right)$(THz)& 9.69\footnotemark[5]& 8.86&
10.1 \\
$\nu _T \left( L \right)$(THz)& 4.19\footnotemark[5]& 3.70&
3.82 \\
$\nu _L \left( K \right)$(THz)& 7.59\footnotemark[5]& 6.87&
7.69 \\
$\nu _{T_1 } \left( K \right)$(THz)& 5.64\footnotemark[5]& 4.80&
5.00 \\
$\nu _{T_2 } \left( K \right)$(THz)& 8.65\footnotemark[5]& 7.76&
8.69 \\
Vacancy:& & &
 \\
$E_v^f $(eV)& 0.68\footnotemark[6]& 0.63&
0.65 \\
Self-interstitial:& & &
 \\
$E_I^f \left( {O_h } \right)$(eV)& $2.73$& 2.10&
2.41 \\
$E_I^f \left( {T_d } \right)$(eV)& $3.08$& 2.55&
2.87 \\
$E_I^f \left( {\left[ {111} \right]\rm{dumbell}} \right)$(eV)& 2.97&
2.48&
2.81 \\
$E_I^f \left( {\left[ {110} \right]\rm{dumbell}} \right)$(eV)& 2.76&
2.12&
2.38 \\
$E_I^f \left( {\left[ {100} \right]\rm{dumbell}} \right)$(eV)&
$2.53$& 2.02&
2.30 \\
Melting temperature:& & &
 \\
$T_m $(K)& 933.6& 593.5$\pm $10&
672.5$\pm $10 \\
\end{tabular}
\label{ap-tab2}
\end{ruledtabular}
\footnotetext[1]{Reference \onlinecite{PR1}.}
\footnotetext[2]{Reference \onlinecite{PR2}.}
\footnotetext[3]{Reference \onlinecite{PR3}.}
\footnotetext[4]{Reference \onlinecite{PR4}.}
\footnotetext[5]{Reference \onlinecite{PR5}.}
\footnotetext[6]{Reference \onlinecite{PR6}.}
\end{table}

Some important properties predicted by VC and GCEAM-VC potentials
are collected in \autoref{ap-tab2}. From \autoref{ap-tab2}, it is
found that the GCEAM-VC improves the overall performance of the VC
potential, especially for high density gradient configurations, such
as self-interstitials. The table clearly demonstrates the success
and improved transferability of GCEAM-VC.

Although GCEAM-VC gives a better result for the melting
temperature than VC, the deviation from the experimental value is
still large. This is due to the fact that the melting process is
associated with long-range interactions, whereas the density
gradient corrections tend to be short-ranged. Therefore the gradient
corrections are not expected to have significant effect on melting
temperature. This is not an intrinsic problem of the GCEAM
model because one could improve the melting temperature by fitting
more accurately the long-range tails of the UDA functions, e.g., $\phi
\left( R \right)$, $\rho ^{\rm{at}}\left( R \right)$, and $F_0
\left( \rho \right)$.

\begin{figure}[htbp]
\includegraphics[width=\linewidth]{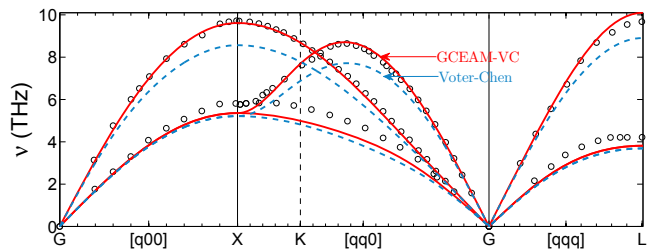}
\caption{The phonon dispersion curves for Al. Red lines are
calculated with GCEAM-VC potential, blue dash lines are calculated
with VC potential, and open circles are experimental data taken from
Ref. \onlinecite{PhDisp}.} \label{ap-fig3}
\end{figure}

In Fig. \ref{ap-fig3}, we compare the phonon dispersions between
GCEAM-VC and VC, against the experimental data. It is found that
GCEAM-VC predicts much better results than VC.

\begin{figure*}[htbp]
\includegraphics[width=\linewidth]{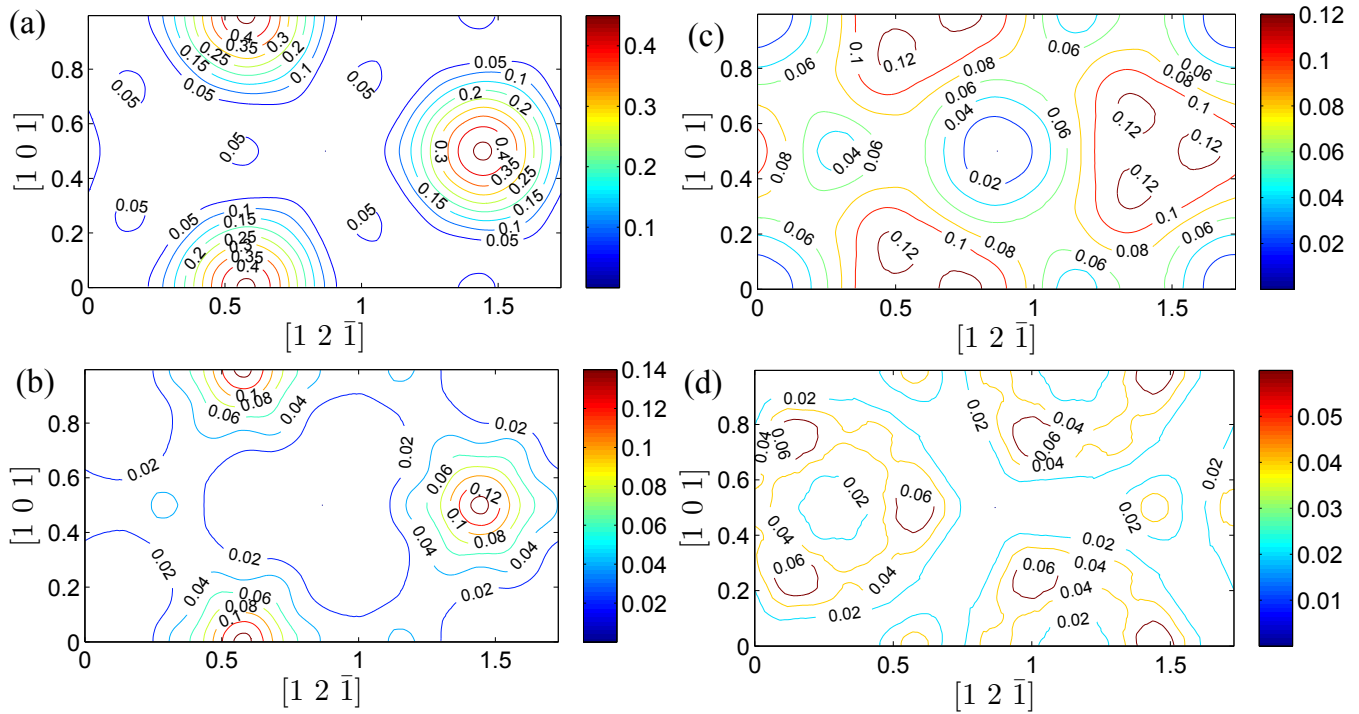}
\caption{The absolute error between the $\gamma$-surface from EAM
potentials and first-principles calculations $\left| E^{\rm
EAM}_{\rm GSF} - E^{\rm VASP}_{\rm GSF}\right|$. In (a), (b), (c)
and (d), the EAM potentials are Ercolessi-Adams, Mishin, VC and
GCEAM-VC, respectively.} \label{ap-fig4}
\end{figure*}

Furthermore, with the help of the sparse-grid method, we calculate
the entire $\gamma $-surface using first-principles VASP method.
From the $\gamma $-surface, one can derive the properties of all
$\{111\}$-type dislocations in Al \cite{GammaSuf}. The absolute
errors between the $\gamma$-surface determined by various EAM
potentials and VASP calculations are shown in Fig. \ref{ap-fig4}.
The projection of the $\gamma$-surface along two special
orientations is plotted in \hyperlink{fig2ab}{Figure 2a and 2b}. It
is found that the GCEAM-VC potential yields the most accurate result
for overall $\gamma $-surface. Here we should emphasize that
GCEAM-VC does not fit \textit{any} stacking fault configurations. In
contrast, Ercolessi-Adams and Mishin \textit{et al.} potentials both
have included $\gamma$-energies in their fitting database, and yet
their results are not as good as GCEAM-VC. This is an important
success of GCEAM potential in terms of transferability. Moreover,
GCEAM-VC potential gives much more accurate stacking fault energy
near the ``run-on'' configuration ($\left( {\frac{\sqrt 3 }{3},0}
\right)$, $\left( {\frac{\sqrt 3 }{3},1} \right)$, and $\left(
{\frac{5\sqrt 3 }{6},\frac{1}{2}} \right)$ in Fig. \ref{ap-fig4}).
These results confirm that indeed the gradient corrections are
crucial for describing high density gradient configurations, such as
the ``run-on'' stacking faults. On the other hand, the errors of
GCEAM-VC appear at configurations where interatomic distance is
greater than that of a perfect lattice. These errors are
not the intrinsic problem of the gradient correction model, but rather due
to the fitting strategy of the Voter-Chen potential.

Finally, it is useful to mention that the force calculation in
GCEAM maintains the comparable numerical efficiency with the standard EAM models.
Thanks to the fact that the modified embedding functions can be
factored by a $\rho$-dependent term and an $s$-dependent term, the
analytical expression of force remains simple - it has
several additional terms that are of similar complexity of that of
standard EAM. To compute these additional terms, the GCEAM needs to
perform extra calculations of which the most time-consuming part is
the second derivatives of the charge density with respect to
distance. By using cubic spline interpolations, these
calculations can be made rather efficient and as a result, the GCEAM
force calculation takes less than twice of the CPU time of the
standard EAM. The code package for calculating the energy and force with GCEAM-VC
potential is available via the World Wide Web\cite{website} or via
e-mail at wugaxp@gmail.com.

\section{Discussion and Conclusion}

Finally, it is instructive to relate the present corrections to other EAM models
\cite{meam, msmeam, cteam}. In the original EAM model, the electron
correlations arising from the inhomogeneous background density are
largely ignored. The goal of the present model is to capture the
missing correlations by taking into consideration of density
gradients. Apart from the inhomogeneity of the density, the correlation
effect also manifests itself in small molecules and clusters - a
well-known fact in quantum chemistry that motivated the development
of GGAs. By introducing a PBE GGA-like correction to the
exchange-correlation part of the embedding energy, the present model
could improve the description of the correlation effect. The
modified EAM (MEAM) and its multistate variant strive to improve the
transferability by making the background density $\bar{\rho}$
angular and reference-state dependent. However, since they are based
on UDA, the MEAM model does not treat the electron correlations
adequately. As a result, it cannot deal with small clusters
accurately as documented in the literature \cite{meam_89}.
The charge transfer EAM (CT-EAM) also
recognizes the importance of the correlation effect. However it
addresses the problem by introducing a reference-state (and its
charge) dependent background density $\bar{\rho}$. Since the present
model considers exchange-correlation energy explicitly, it can
achieve the same goal of CT-EAM with a simpler function form.
Moreover, one could incorporate MEAM and its variants into
the present model by making the background density $\bar{\rho}$ in
Eq. (\ref{eq1}) as angular, reference-state and/or charge-dependent
if so desired.


In conclusion, we have performed detailed EAM and first-principles
calculations of Al for elastic deformation and generalized stacking
fault energy. We find that although EAM models reproduce well the
first-principles results for most cases, they tend to fail when the
electron density gradients become substantial. We attribute the
failures of EAM to the violation of UDA underlying the existing EAM
models. To remedy the deficiency of UDA, we propose an improved EAM
model which considers explicitly the gradient corrections to the
embedding function in terms of the exchange-correlation energy and
the kinetic energy. We show that the gradient corrected model can
significantly improve the transferability of EAM, and represents a new
direction for developing more transferable EAM potentials.

\acknowledgements{
The research at California State University Northridge was supported
in part by DOE grant DE-FC02-06ER25791 and NSF grant
DMR-0611562. CJGC's work was funded by an NSF CAREER award.
We thank Art Voter for valuable comments on Voter-Chen potential.}

\end{document}